\documentclass[letterpaper, 10 pt, conference]{ieeeconf}

\IEEEoverridecommandlockouts
\usepackage[english]{babel}
\usepackage{amsmath}
\usepackage{mathtools}
\usepackage{amssymb,bm}
\IfFileExists{bbm.sty}{\usepackage{bbm}}{}

\usepackage{graphicx}
\usepackage{tikz}
\usepackage{pgfplots}
\pgfplotsset{compat=1.18}
\usepackage{tkz-graph}

\usetikzlibrary{
  arrows,
  arrows.meta,
  positioning,
  shapes,
  shapes.geometric,
  shapes.misc,        % <--- ADD THIS
  calc,
  fit,
  backgrounds,
  decorations.pathreplacing,
  patterns.meta
}

\usepackage{booktabs}
\usepackage{multirow}
\usepackage{tabularx}

\let\labelindent\relax
\usepackage{enumitem}

\usepackage{subcaption}
\usepackage[noadjust]{cite}

\makeatletter
\@ifundefined{NAT@parse}{}{\let\NAT@parse\undefined}
\makeatother

\providecommand{\citep}[1]{\cite{#1}}
\providecommand{\citet}[1]{\cite{#1}}

\providecommand{\citeauthor}[1]{\cite{#1}}
\providecommand{\citeyear}[1]{\cite{#1}}

\usepackage[hidelinks]{hyperref}
\usepackage[capitalize]{cleveref}

\definecolor{myblue}{HTML}{2166AC}     % trapping
\definecolor{mygreen}{HTML}{1B7837}    % transient
\definecolor{myorange}{HTML}{E08214}   % marginal

\definecolor{myred}{HTML}{B2182B}      % persistent dynamics: periodic orbits
\definecolor{mytarget}{HTML}{1A1A1A}   % static geometry: targets

\newtheorem{theorem}{Theorem}
\newtheorem{definition}{Definition}
\newtheorem{proposition}{Proposition}

\newtheorem{lemma}{Lemma}

\newtheorem{standassumption}{Standing Assumption}
\crefname{standassumption}{}{}
\Crefname{standassumption}{}{}

\crefformat{standassumption}{#2#1#3}
\Crefformat{standassumption}{#2#1#3}
\crefmultiformat{standassumption}{#2#1#3}{ and #2#1#3}{, #2#1#3}{ and #2#1#3}
\Crefmultiformat{standassumption}{#2#1#3}{ and #2#1#3}{, #2#1#3}{ and #2#1#3}

\newtheorem{remark}{Remark}

\newtheorem{example}{Example}

\makeatletter
\def\@begintheorem#1#2{%
  \trivlist\item[]\relax
  \textit{\indent #1\ #2:}\ \ignorespaces}
\def\@opargbegintheorem#1#2#3{%
  \trivlist\item[]\relax
  \textit{\indent #1\ #2\ (#3):}\ \ignorespaces}
\makeatother
\makeatletter
\newif\ifrp@short
\DeclareRobustCommand{\rp@fmt}[2]{\ifrp@short(#2)\else#1(#2)\fi}
\newcommand{\partref}[1]{{\rp@shorttrue\ref{#1}}}
\def\rp@gettype[#1][#2]#3\@nil{\def\rp@type{#1}}
\setlist[enumerate,1]{label=(\roman*), ref=\rp@fmt{\rp@parent}{\roman*}, wide}
\SetEnumitemKey{alph}{label=(\alph*), ref=\rp@fmt{\rp@parent}{\alph*}}
\AddToHook{env/enumerate/begin}{%
  \ifnum\@enumdepth=\z@
    \protected@edef\rp@parent{\@currentlabel}%
    \expandafter\rp@gettype\cref@currentlabel\@nil
    \crefalias{enumi}{\rp@type}%
  \fi}
\makeatother

\makeatletter
\def\@begintheorem#1#2{%
  \trivlist\item[]\relax
  \textit{\indent \textbf{#1}\ #2:}\ \ignorespaces}
\def\@opargbegintheorem#1#2#3{%
  \trivlist\item[]\relax
  \textit{\indent \textbf{#1}\ #2\ (#3):}\ \ignorespaces}
\makeatother
\usepackage{stfloats}
\usepackage{balance}

\title{\LARGE \bf
Robust Action Clusters and Cascades in Opinion Dynamics with Heterogeneous Action Thresholds
}

\author{Hassan~Munif$^{1}$, Anthony~Couthures$^{2}$, Vineeth~S.~Varma$^{1}$, Samson~Lasaulce$^{1}$, and Irinel-Constantin~Mor\u{a}rescu$^{1}$
\thanks{This work was supported by the CNRS MITI project BLESS. The authors used Claude Opus 5.5 for proofreading and numerical illustrations.
}%
\thanks{$^{1}$Universit\'e de Lorraine, CNRS, CRAN, F-54000 Nancy, France.
        Corresponding author: Hassan~Munif.\protect\newline {\tt\small hassan.mohamad@univ-lorraine.fr}}%
\thanks{$^{2}$UCLouvain, ICTEAM Institute, Louvain-la-Neuve, 1348, Belgium.}%
}

\begin{document}

\maketitle
\thispagestyle{empty}
\pagestyle{empty}
\raggedbottom

\begin{abstract}
We study the preservation and propagation of public actions in networks whose agents hold continuous private opinions. Agents update their opinions from their neighbors' observed actions and express them through action rules whose thresholds differ across agents. We characterize groups of agents that preserve a given common action regardless of the outsiders' actions. We call such a group, paired with that action, a robust action cluster. We show that a group is a robust action cluster for that action if and only if each member places enough of its influence weight on the group, where the required influence depends on the member's own action rule. We further give a sufficient condition under which an agent outside the cluster permanently adopts the cluster's action, and we iterate it to obtain a cascade of agents that are guaranteed to adopt that action. Finally, numerical examples show that replacing heterogeneous thresholds by their mean can change both the propagation of an action and the presence of robust action clusters.
\end{abstract}
  % Finally, numerical examples on a small network and a square lattice show that heterogeneous action rules can enable permanent adoption of an action when a common homogeneous action rule with their mean thresholds does not, and that a fixed action profile need not contain a robust action cluster.

\section{Introduction}
\label{sec:intro}

People often hold a nuanced private opinion but express it through a coarse public choice: a vote, a rating, or one of a few stated positions. Classical averaging models describe social influence through the exchange of opinions~\cite{degroot1974reaching}. Continuous-opinion discrete-action (CODA) models distinguish these private opinions from the public actions through which agents influence one another~\cite{martins2008continuous,chowdhury2016continuous}. Two people can express the same position despite holding different opinions or using different rules to translate opinions into actions.
A group defined by its public behavior can therefore be heterogeneous in both respects.

This distinction raises two questions. When can a group preserve a common action despite influence from the rest of the network? When does that action spread to others? Threshold models of collective behavior show that aggregate outcomes can depend on how individual response thresholds are distributed, not only on their mean~\cite{granovetter1978threshold,watts2002simple,rossi2019threshold}.
In our setting, by contrast, each action is determined by a private opinion that retains memory of past interactions. No agent is stubborn: a group can preserve an action only through the opinion dynamics that all agents follow.

\textit{Related work.} The binary CODA model of~\cite{chowdhury2016continuous} studies preservation and propagation through majority conditions. Its state-dependent gain and history-dependent tie rule differ from the dynamics considered here. Related stochastic models couple multilevel opinions with binary actions~\cite{varma2017modeling}. Quantized behaviors are also analyzed in continuous time~\cite{ceragioli2018consensus,ceragioli2025disruptive} and coupled with external dynamics~\cite{couthures2024analysis}. Complementary models study discrepancies between private and expressed opinions caused by social conformity~\cite{ye2019influence} and the coevolution of binary actions and continuous opinions on two-layer networks~\cite{aghbolagh2023coevolutionary}. Strategic communication on influence networks can also generate heterogeneous quantized communication rules~\cite{munif2026strategic}.
Here, the action rules are prescribed and share a common finite action set.

Beyond binary actions, finite verbal scales~\cite{urbig2003attitude} and a middle option~\cite{martins2009middle} have been studied by simulation. Multistage contagion models further show that additional states can change cascade dynamics~\cite{melnik2013multistage}. In threshold dynamics, cohesive groups can obstruct contagion~\cite{morris2000contagion}, while robust coordination under asynchronous binary updates is studied in~\cite{arditti2024robust}. These works motivate examining how network cohesion and individual action rules jointly determine action preservation and propagation.

We study this interplay on weighted directed networks with inertial private opinions and
heterogeneous thresholds on a common finite action set. For intermediate actions, robustness requires a two-sided cohesion condition that protects against influence toward both lower and higher actions.
Actions can switch indefinitely, even with two agents (\cref{ex:two_cycle}), so we study permanent adoption without assuming convergence of the entire network. Our contributions are threefold. First, we give a necessary and sufficient condition for a group to be a robust action cluster. It compares the influence weight each member places on the group with an agent-dependent influence threshold set by that member's action rule. Thus, it separates the group's cohesion from the individual rules.
Second, a strict version of this condition guarantees that an agent placing enough influence weight on a robust action cluster adopts its action permanently, from any initial opinion. Applying it repeatedly yields a finite cascade of guaranteed adopters and
a sufficient condition for permanent consensus on that action. The cascade identifies all guaranteed adopters for two actions, but provides only a sufficient condition for larger action sets. Third, two numerical examples illustrate that mean thresholds alone determine neither
the propagation of an intermediate action from a robust action cluster
nor the presence of robust action clusters.
On a lattice, heterogeneous rules produce robust action clusters.
The corresponding homogeneous rule, which gives every agent the mean thresholds, yields none in its final action profile, although the action profile remains fixed in both runs.

The model and its general properties are introduced in \cref{sec:model,sec:general}; \cref{sec:retention,sec:propagation} establish preservation and adoption conditions. \Cref{sec:numerics} illustrates permanent adoption from a robust seed and distinguishes action profiles that remain fixed from \mbox{robust action clusters.}

\textit{Notation.} Vectors and matrices are bold (e.g., $\bm{v} \in \mathbb{R}^n$, $\bm{W} \in \mathbb{R}^{n \times m}$), with $v_i$ and $W_{ij}$ denoting their respective entries. Vector inequalities are componentwise: $\bm x\leq\bm y$ means $x_i\leq y_i$ for every $i$.
The vector $\bm{1}$ has all entries equal to one, and $\bm{I}$ is the identity matrix.
We write $|\mathcal{S}|$ for set cardinality, and $\mathbb{R}_{+}$ for the set of nonnegative reals.
We denote the interior and closure of a set $\mathcal{C}$ \mbox{by $\mathcal{C}^{\circ}$ and $\overline{\mathcal{C}}$, respectively.}

\section{Model}
\label{sec:model}

Consider $N$ agents indexed by $\mathcal{V}\coloneqq\{1,\dots,N\}$. Their interactions are described by a weighted directed graph $\mathcal{G}=(\mathcal{V},\mathcal{E},\bm{W})$, where $\bm{W}=[W_{ij}]\in\mathbb{R}_{+}^{N\times N}$. The entry $W_{ij}$ is the \emph{influence weight}
of agent $j$ on agent $i$. A link $(j,i)\in\mathcal{E}$ is present if and only if $W_{ij}>0$. Thus $\mathcal{N}_i\coloneqq\{j\in\mathcal{V}:(j,i)\in\mathcal{E}\}$ is the set of agents observed by $i$. We normalize each agent's influence weights and assume the following properties for the interaction graph.
\begin{standassumption}
  \label[standassumption]{asm:graph}
  The matrix $\bm{W}$ is time-invariant, nonnegative, and row-stochastic ($\bm{W}\bm{1}=\bm{1}$), \mbox{with zero diagonal.}
\end{standassumption}
For a group $\mathcal{R}\subseteq\mathcal{V}$, agent $i$ places influence weight $w_{i}(\mathcal{R})\coloneqq\sum_{j\in\mathcal{R}}W_{ij}$ on $\mathcal{R}$. By \cref{asm:graph}, it is the fraction of $i$'s total influence weight, so $w_i(\mathcal{V}\setminus\mathcal{R})=1-w_i(\mathcal{R})$.

\subsection{Private Opinions and Public Actions}
\label{subsec:quantizers}

At each time $t\in\{0,1,2,\dots\}$, agent $i$ holds a private opinion $x_i(t)\in[-1,1]$ and broadcasts a public action. Agents observe only their neighbors' actions.

All agents choose from a common ordered action set $\mathcal{A}\coloneqq\{a_1<\dots<a_L\}\subset[-1,1]$, with $L\geq2$. The actions strictly between $a_1$ and $a_L$ are the \emph{intermediate actions}.
At time $t$, agent $i$ broadcasts the action $q_i(x_i(t))$, a quantized version of its opinion. The map $q_i$, called the \emph{action rule} of agent $i$, assigns an action to each opinion through agent-dependent thresholds.
It is defined as follows.
\begin{definition}[Action rule]
  \label{def:quantizer}
  For each $i\in\mathcal{V}$, let $-1<\theta_{i,1}<\dots<\theta_{i,L-1}<1$ be the thresholds of agent $i$, with the conventions $\theta_{i,0}\coloneqq-\infty$ and $\theta_{i,L}\coloneqq+\infty$. The action rule $q_i:\mathbb{R}\to\mathcal{A}$ of agent $i$ is
  \begin{equation}
    \label{eq:quantizer}
    q_i(x)\coloneqq a_\ell\quad\text{if}\quad x\in[\theta_{i,\ell-1},\theta_{i,\ell}),\qquad \ell=1,\dots,L.
  \end{equation}
\end{definition}

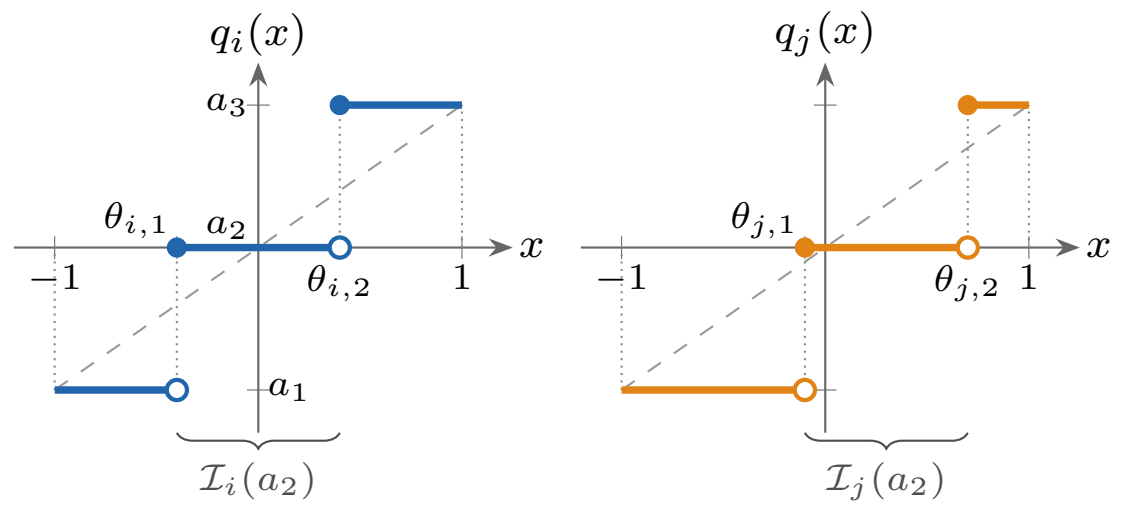
\begin{figure}[t]
  \centering
  % figure1_Tikz.tex -- quantizers of two agents on the opinion range [-1,1], centred axes.
% Data from Remark rem:neutral: A={-1,0,1}, I_i(0)=[-0.4,0.4), I_j(0)=[-0.1,0.7).
% The a_2 = 0 step lies on the x-axis; a_1 = -1 below it, a_3 = 1 above it.
\begin{tikzpicture}[
    stp/.style={line width=1.4pt},
    cl/.style={circle, fill, inner sep=0pt, minimum size=3.8pt},
    op/.style={circle, draw, fill=white, line width=0.8pt, inner sep=0pt, minimum size=3.8pt},
    guidedot/.style={densely dotted, black!45},
    lvl/.style={font=\scriptsize, inner sep=1.5pt},
  ]
  \pgfplotsset{
    qaxis/.style={
      scale only axis, width=3.3cm, height=2.45cm,
      xmin=-1.2, xmax=1.25, ymin=-1.3, ymax=1.3,
      axis lines=middle, axis line style={black!60, -{Stealth[length=1.6mm]}},
      ytick={-1,1}, yticklabels={},
      tick style={black!60}, tick label style={font=\scriptsize, fill=white, inner sep=1pt},
      xlabel={$x$}, label style={font=\footnotesize},
      xlabel style={at={(ticklabel* cs:1)}, anchor=west, inner sep=1pt},
      ylabel style={at={(ticklabel* cs:1)}, anchor=south, inner sep=1pt},
      clip=false,
    },
  }
  % #1 theta1, #2 theta2, #3 colour, #4 agent
  \newcommand{\staircase}[4]{%
    \addplot[dashed, black!40, thin, domain=-1:1, samples=2] {x};
    \draw[guidedot] (axis cs:-1,-1) -- (axis cs:-1,0);
    \draw[guidedot] (axis cs:#1,-1) -- (axis cs:#1,0);
    \draw[guidedot] (axis cs:#2,0)  -- (axis cs:#2,1);
    \draw[guidedot] (axis cs:1,0)   -- (axis cs:1,1);
    \draw[stp, #3] (axis cs:-1,-1) -- (axis cs:#1,-1);
    \draw[stp, #3] (axis cs:#1,0)  -- (axis cs:#2,0);
    \draw[stp, #3] (axis cs:#2,1)  -- (axis cs:1,1);
    \node[op, draw=#3] at (axis cs:#1,-1) {};
    \node[cl, #3]      at (axis cs:#1,0) {};
    \node[op, draw=#3] at (axis cs:#2,0) {};
    \node[cl, #3]      at (axis cs:#2,1) {};
    % theta_{.,1} labelled above the x-axis, left of its guide (keeps it clear of the q-axis)
    \node[anchor=south east, font=\scriptsize, inner sep=1pt] at (axis cs:#1,0.02) {$\theta_{#4,1}$};
    % brace marking the quantization interval of a_2
    \draw[decorate, decoration={brace, mirror, amplitude=3pt}, black!70]
      (axis cs:#1,-1.3) -- node[below=2pt, font=\scriptsize] {$\mathcal{I}_{#4}(a_2)$} (axis cs:#2,-1.3);
  }

  \begin{axis}[qaxis, name=left, ylabel={$q_i(x)$},
      xtick={-1,-0.4,0.4,1}, xticklabels={$-1$,{},$\theta_{i,2}$,$1$}]
    \staircase{-0.4}{0.4}{myblue}{i}
    % action levels: a_3 above the x-axis, a_2 on it, a_1 below it
    \node[lvl, anchor=east]       at (axis cs:0,1)  {$a_3$};
    \node[lvl, anchor=south east] at (axis cs:0,0)  {$a_2$};
    \node[lvl, anchor=west]       at (axis cs:0,-1) {$a_1$};
  \end{axis}

  \begin{axis}[qaxis, at={($(left.east)+(0.45cm,0)$)}, anchor=west, ylabel={$q_j(x)$},
      xtick={-1,-0.1,0.7,1}, xticklabels={$-1$,{},$\theta_{j,2}$,$1$}]
    \staircase{-0.1}{0.7}{myorange}{j}
  \end{axis}
\end{tikzpicture}
  \caption{Action rules of two agents with $\mathcal{A}=\{-1,0,1\}$, $\mathcal{I}_i(a_2)=[-0.4,0.4)$, and $\mathcal{I}_j(a_2)=[-0.1,0.7)$, on the opinion range $[-1,1]$. Filled and open dots show right-continuity. Each step meets the identity (dashed) at its own action level, as required by \cref{asm:self_consistent}.}
  \label{fig:quantizers}\vspace{-3mm}
\end{figure}

By construction, each action rule $q_i$ is nondecreasing and right-continuous: at a threshold, the agent takes the higher of the two adjacent actions (\cref{fig:quantizers}).

The opinions at which agent $i\in\mathcal{V}$ takes action $a\in\mathcal{A}$ form the \emph{action interval}
\begin{equation}
  \label{eq:action_interval}
  \mathcal{I}_{i}(a)\coloneqq\{x\in\mathbb{R}: q_i(x)=a\}=[\underline{\theta}_{i}(a), \overline{\theta}_{i}(a)),
\end{equation}
with lower and upper thresholds $\underline{\theta}_{i}(a_\ell)\coloneqq\theta_{i,\ell-1}$ and $\overline{\theta}_{i}(a_\ell)\coloneqq\theta_{i,\ell}$. Only $\underline{\theta}_{i}(a_1)$ and $\overline{\theta}_{i}(a_L)$ are infinite.

We require self-consistency: each action is a fixed point of \mbox{every action rule}~\cite{ceragioli2018consensus}.

\begin{standassumption}
  \label[standassumption]{asm:self_consistent}
  For all $i\in\mathcal{V}$ and $a\in\mathcal{A}$, $q_i(a)=a$; equivalently, $a\in\mathcal{I}_i(a)$.
\end{standassumption}

In terms of the thresholds, this means that $a_\ell<\theta_{i,\ell}\leq a_{\ell+1}$ for every $i\in\mathcal{V}$ and $\ell=1,\dots,L-1$.

\subsection{Opinion Dynamics}
\label{subsec:dynamics}

The observed actions in turn drive the evolution of private opinions. The social input to agent $i$ is the weighted average of its neighbors' actions, $u_i(t)\coloneqq\sum_{j\in\mathcal{N}_i}W_{ij}\,q_j(x_j(t))
$. It depends on the neighbors' opinions only through their actions. The opinion of agent $i$ evolves according to the discrete-time dynamics
\begin{equation}
  \label{eq:dynamics}
  x_i(t+1)=\alpha_i x_i(t)+(1-\alpha_i)\,u_i(t),
\end{equation}
where $\alpha_i\in(0,1)$ is the \emph{inertia} (self-persistence) parameter of agent $i$, determining the weight assigned to its current opinion. Since $W_{ii}=0$, the social input
excludes the agent's own action, and its own opinion enters only through inertia.

We write $\bm{x}(t)\coloneqq(x_1(t),\dots,x_N(t))^\top$ and $\bm{q}(\bm{x})\coloneqq(q_1(x_1),\dots,q_N(x_N))^\top$. With $\bm{D}_{\alpha}\coloneqq\operatorname{diag}(\alpha_1,\dots,\alpha_N)$, the dynamics take the vector form
\begin{equation}
  \label{eq:dynamics_vector}
  \bm{x}(t+1)=\bm{D}_{\alpha}\bm{x}(t)+(\bm{I}-\bm{D}_{\alpha})\bm{W}\bm{q}(\bm{x}(t))\eqqcolon\bm{f}(\bm{x}(t)).
\end{equation}
The public \emph{action profile} at time $t$ is thus $\bm{q}(\bm{x}(t))$, and the state space is $\mathcal{X}\coloneqq[-1,1]^N$.
The rest of the paper analyzes this model and characterizes when a group of agents
preserves or adopts a common action permanently.

\section{General Properties of the Opinion Dynamics}
\label{sec:general}

We first establish basic structural properties of the map $\bm{f}$ on the state space $\mathcal{X}$.

\begin{lemma}[{Well-posedness and monotonicity}]
  \label{lem:basic}
  We have:
  \begin{enumerate}
    \item\label{lem:basic:i} $\bm{f}(\mathcal{X})\subseteq\mathcal{X}$.
    \item\label{lem:basic:ii} $\bm{f}$ is monotone: $\bm{x}\leq\bm{y}$ implies $\bm{f}(\bm{x})\leq\bm{f}(\bm{y})$.
  \end{enumerate}
\end{lemma}

\begin{proof}
  \partref{lem:basic:i} Since $\mathcal{A}\subset[-1,1]$ and $\bm{W}$ is row-stochastic, $u_i\in[-1,1]$, and each update \eqref{eq:dynamics} is a convex combination of $x_i$ and $u_i$.
  \partref{lem:basic:ii} The matrices $\bm{D}_{\alpha}$ and $(\bm{I}-\bm{D}_{\alpha})\bm{W}$ are nonnegative, and each $q_i$ is nondecreasing.
\end{proof}

Next, we show that if all agents share the same action, they preserve it for all time. For any group $\mathcal{R}\subseteq\mathcal{V}$ and any action $a\in\mathcal{A}$, let
\(
\mathcal{X}_{a}(\mathcal{R})\coloneqq\{\,\bm{x}\in\mathcal{X} : q_i(x_i)=a\ \text{for all}\ i\in\mathcal{R}\,\}
\)
be the states in which every member of $\mathcal{R}$ holds action $a$, with the other agents unconstrained.

\begin{lemma}[Global action consensus]
  \label{lem:globalcons}
  For every $a\in\mathcal{A}$, the consensus state $a\bm{1}$ is a fixed point of $\bm{f}$, and $\bm{x}\in\mathcal{X}_a(\mathcal{V})$ implies $\bm{f}(\bm{x})\in\mathcal{X}_a(\mathcal{V})$.
\end{lemma}

\begin{proof}
  By \cref{asm:self_consistent}, $a\in\mathcal{I}_i(a)$ for all $i\in\mathcal{V}$, so $\bm{q}(a\bm{1})=a\bm{1}$ and $\bm{f}(a\bm{1})=a\bm{1}$ by row-stochasticity of $\bm{W}$. More generally, if $\bm{x}\in\mathcal{X}_a(\mathcal{V})$, then $u_i=a$ for every $i\in\mathcal{V}$, so \eqref{eq:dynamics} gives $f_i(\bm{x})=\alpha_i x_i+(1-\alpha_i)a$. Since $x_i,a\in\mathcal{I}_i(a)$, their convex combination lies in $\mathcal{I}_i(a)$, and \cref{lem:basic}\partref{lem:basic:i} ensures $\bm{f}(\bm{x})\in\mathcal{X}$, so $\bm{f}(\bm{x})\in\mathcal{X}_a(\mathcal{V})$.
\end{proof}

By \cref{lem:basic}\partref{lem:basic:ii}, ordered initial states give ordered trajectories.
Monotonicity, however, does not imply convergence, since it still allows persistent switching. As in synchronous threshold dynamics~\cite{goles1980periodic}, agents can switch actions periodically, even on a connected graph.

The following example shows that two agents can enter a period-two orbit, preventing opinion convergence.
\begin{example}[Persistent switching]
  \label{ex:two_cycle}
  Consider two agents with action set $\mathcal{A}=\{-a,a\}$, where $0<a\leq1$, and dynamics specified by
  \begin{equation*}
    \label{eq:swap_network}
    \bm W=\begin{bmatrix}0&1\\1&0\end{bmatrix},
    \qquad
    q_i(x)=\begin{cases}-a,&x<0,\\ a,&x\geq0.\end{cases}
  \end{equation*}
  For arbitrary $\alpha_1,\alpha_2\in(0,1)$, define
  $
    r_i\coloneqq a\frac{1-\alpha_i}{1+\alpha_i}>0.
  $

  The opinion update maps $(r_1,-r_2)$ to $(-r_1,r_2)$ and back.
  Both states avoid the threshold because $r_i>0$, so the agents' actions switch indefinitely.
\end{example}

Since the action profile need not converge, we study which agents hold a given action permanently, rather than convergence of the whole network. \Cref{sec:retention} provides a necessary and sufficient condition under which a group of agents keeps a common action independently of what the rest of the network
does. This generalizes \cref{lem:globalcons} from $\mathcal{V}$ to any $\mathcal{R}\subseteq\mathcal{V}$. \Cref{sec:propagation} then provides sufficient conditions under which that action propagates permanently to agents outside the group.

\section{Robust Action Clusters}
\label{sec:retention}

Agents holding a common action reinforce it through mutual influence, while the other agents may hold different actions.
In this section, we investigate when such a group preserves its action regardless of the rest of the network.
We start with the central definition of the paper, which concerns a group of agents paired with a public action.

\begin{definition}[Robust action cluster]
  \label{def:robust_cluster}
  Let $a\in\mathcal{A}$ and $\emptyset\neq\mathcal{R}\subseteq\mathcal{V}$.
  $\mathcal{R}$ is a \emph{robust action cluster for action $a$} if
  \begin{equation}
    \label{eq:def_robust_cluster}
    \bm{x}\in\mathcal{X}_{a}(\mathcal{R})\implies\bm{f}(\bm{x})\in\mathcal{X}_{a}(\mathcal{R}),
  \end{equation}
  i.e., $\mathcal{X}_{a}(\mathcal{R})$ is forward invariant under the dynamics \eqref{eq:dynamics_vector}.
  For shorthand, we write ``$(\mathcal{R},a)$ is robust''. The agents in $\mathcal{V}\setminus\mathcal{R}$ are the \emph{outsiders} of $\mathcal{R}$.
\end{definition}

By \cref{lem:globalcons}, $(\mathcal{V},a)$ is always robust. In general,
a robust action cluster preserves its action from every initial state in $\mathcal{X}_a(\mathcal{R})$, whatever the actions of its outsiders.
Our first objective is to characterize the robust action clusters for each action $a\in\mathcal{A}$. The second, addressed in \cref{sec:propagation}, is to determine when their action spreads to outsiders.

From the dynamics in \eqref{eq:dynamics}, each updated opinion is a convex combination of the current opinion and the social input. Hence, an agent whose opinion and input both lie in $\mathcal{I}_i(a)$ keeps action $a$.
We therefore characterize robustness of $(\mathcal{R},a)$ by bounding the social inputs that \mbox{an agent can receive.}

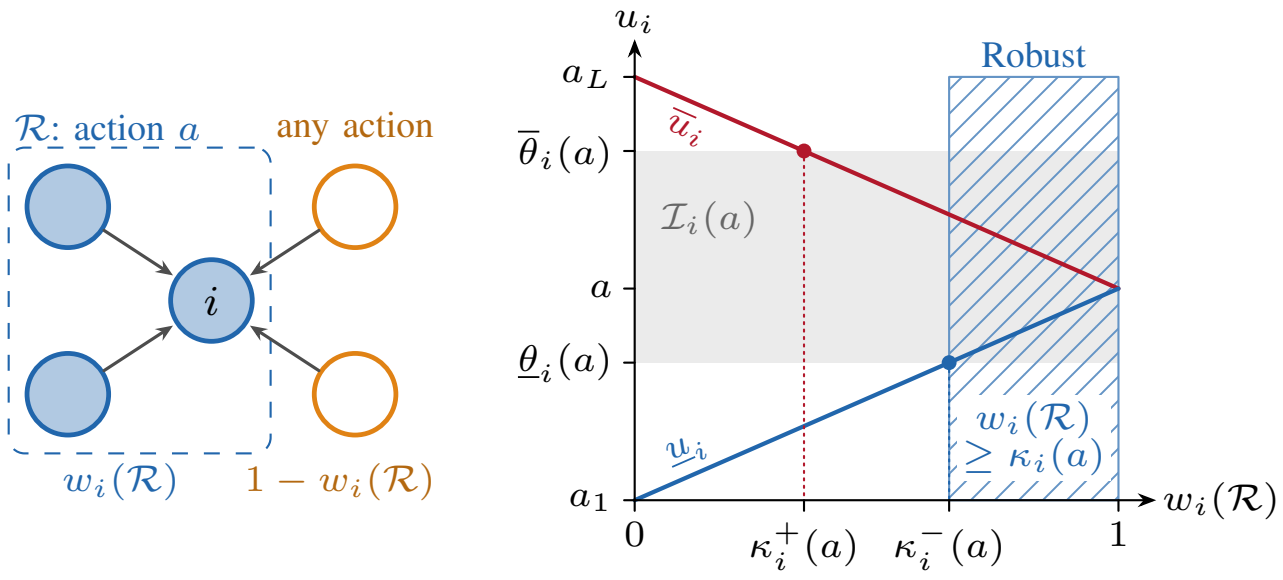
\begin{figure}[t]
  \centering
  \begin{subfigure}[b]{0.36\columnwidth}
    \centering
    % Fig. influence threshold (a): influence on agent i. Arrows point in the
% direction of influence: weight w_i(R) from its group R (holding a),
% 1-w_i(R) from the other agents (any action).
\begin{tikzpicture}[
    ag/.style={circle, draw=black!70, line width=0.6pt, minimum size=5.4mm, inner sep=0pt, font=\footnotesize},
    sd/.style={ag, draw=myblue, fill=myblue!35, line width=0.9pt},
    od/.style={ag, draw=myorange, fill=white, line width=0.9pt},
    ed/.style={line width=0.6pt, black!70, -{Stealth[length=3.2pt,width=2.6pt]}, shorten >=0.5pt},
    line cap=round, line join=round,
  ]
  \node[sd] (i)  at (0,0) {$i$};
  \node[sd] (r1) at (-0.95,0.62) {};
  \node[sd] (r2) at (-0.95,-0.62) {};
  \node[od] (o1) at (0.95,0.62) {};
  \node[od] (o2) at (0.95,-0.62) {};
  \draw[ed] (r1) -- (i);
  \draw[ed] (r2) -- (i);
  \draw[ed] (o1) -- (i);
  \draw[ed] (o2) -- (i);
  \begin{scope}[on background layer]
    \node[draw=myblue, dashed, rounded corners=3pt, fit=(r1)(r2)(i), inner sep=3pt] (box) {};
  \end{scope}
  \node[font=\scriptsize, text=myblue, anchor=south west, inner sep=1pt] at (box.north west) {$\mathcal{R}$: action $a$};
  \node[font=\scriptsize, myorange!80!black, inner sep=1pt] at (0.95,1.14) {any action};
  \node[font=\scriptsize, myblue] at (-0.6,-1.2) {$w_i(\mathcal{R})$};
  \node[font=\scriptsize, myorange!80!black] at (0.85,-1.2) {$1-w_i(\mathcal{R})$};
  % Invisible extent: centers (a) vertically against the plot in (b).
  \path (0,-1.85) (0,1.75);
\end{tikzpicture}%
    \caption{}
    \label{fig:influence_threshold_agent}
  \end{subfigure}%
  \hfill
  \begin{subfigure}[b]{0.62\columnwidth}
    \centering
    % Fig. influence threshold (b): worst-case input bounds versus w_i(R).
% Agent i keeps action a for w_i(R) >= kappa_i(a) (hatched).
\begin{tikzpicture}[font=\scriptsize, line cap=round, line join=round]
  \pgfmathsetmacro{\W}{3.2}
  \pgfmathsetmacro{\H}{2.8}
  \pgfmathsetmacro{\ya}{0.5*\H}
  \pgfmathsetmacro{\kp}{0.35} % kappa_i^+(a)
  \pgfmathsetmacro{\km}{0.65} % kappa_i^-(a) = kappa_i(a)
  \pgfmathsetmacro{\xp}{\kp*\W}
  \pgfmathsetmacro{\xm}{\km*\W}
  \pgfmathsetmacro{\yup}{\ya+(1-\kp)*(\H-\ya)}   % theta-bar
  \pgfmathsetmacro{\ylo}{\ya-(1-\km)*\ya}        % theta-underbar
  % Action interval I_i(a) and the robust range w_i(R) >= kappa_i(a) (hatched).
  \fill[black!8] (0,\ylo) rectangle (\W,\yup);
  \fill[pattern={Lines[angle=45, distance=3.2pt, line width=0.35pt]}, pattern color=myblue!70] (\xm,0) rectangle (\W,\H);
  \draw[myblue!70, line width=0.4pt] (\xm,0) rectangle (\W,\H);
  \node[myblue, above, inner sep=1.5pt] at ({0.5*(\xm+\W)},\H) {Robust};
  \node[black!60, anchor=west] at (0.05,{0.5*(\ya+\yup)}) {$\mathcal{I}_i(a)$};
  \node[myblue, align=center, fill=white, inner sep=1pt] at ({0.5*(\xm+\W)},{0.45*\ylo}) {$w_i(\mathcal{R})$\\[-1pt]$\geq\kappa_i(a)$};
  % Bounds of the worst-case input interval.
  \draw[myred, thick] (0,\H) -- (\W,\ya) node[pos=0.13, below, sloped, inner sep=1.5pt] {$\overline{u}_i$};
  \draw[myblue, thick] (0,0) -- (\W,\ya) node[pos=0.13, above, sloped, inner sep=1.5pt] {$\underline{u}_i$};
  % Crossings of the thresholds.
  \draw[myred, densely dotted] (\xp,\yup) -- (\xp,0);
  \draw[myblue, densely dotted] (\xm,\ylo) -- (\xm,0);
  \fill[myred] (\xp,\yup) circle (1.4pt);
  \fill[myblue] (\xm,\ylo) circle (1.4pt);
  % Axes.
  \draw[-{Stealth[length=3.5pt]}] (0,0) -- (\W+0.25,0) node[right, inner sep=1pt] {$w_i(\mathcal{R})$};
  \draw[-{Stealth[length=3.5pt]}] (0,0) -- (0,\H+0.25) node[above, inner sep=1pt] {$u_i$};
  \foreach \x/\l in {0/{$0$}, \xp/{$\kappa_i^+(a)$}, \xm/{$\kappa_i^-(a)$}, \W/{$1$}}
    \draw (\x,0) -- (\x,-0.07) node[below, inner sep=2pt, anchor=north] {\l};
  \foreach \y/\l in {0/{$a_1$}, \ylo/{$\underline{\theta}_i(a)$}, \ya/{$a$}, \yup/{$\overline{\theta}_i(a)$}, \H/{$a_L$}}
    \draw (0,\y) -- (-0.07,\y) node[left, inner sep=2pt] {\l};
\end{tikzpicture}%
    \caption{}
    \label{fig:influence_threshold_bounds}
  \end{subfigure}
  \caption{Influence threshold of agent $i\in\mathcal{R}$ for action $a$. (a) Agent $i$ places influence weight $w_i(\mathcal{R})$ on its group, which holds $a$, and ${1-w_i(\mathcal{R})}$ on the other agents, which may hold any action. (b) Worst-case input bounds \eqref{eq:input_bounds} versus $w_i(\mathcal{R})$. They cross the thresholds of $\mathcal{I}_i(a)$ (gray) at $\kappa_i^+(a)$ and $\kappa_i^-(a)$, and both stay between them exactly when $w_i(\mathcal{R})\geq\kappa_i(a)$ (hatched); here, $\kappa_i(a)=\kappa_i^-(a)$.}
  \label{fig:influence_threshold}\vspace{-3mm}
\end{figure}

If $\bm{x}(t)\in\mathcal{X}_a(\mathcal{R})$, splitting the social input
between $\mathcal{R}$ and its outsiders gives $u_i(t)=a\,w_i(\mathcal{R})+\sum_{j\notin\mathcal{R}}W_{ij}\,q_j(x_j(t))$. Since every action lies in $[a_1,a_L]$ and the weights $W_{ij}$ are nonnegative, each agent's social input lies in the worst-case
input interval $\mathcal{U}_{i}(\mathcal{R},a)\coloneqq [\underline{u}_{i}(\mathcal{R},a),\overline{u}_{i}(\mathcal{R},a)]$, where
\begin{equation}
  \label{eq:input_bounds}
  \begin{aligned}
    \underline{u}_{i}(\mathcal{R},a) & \coloneqq a\,w_i(\mathcal{R})+(1-w_i(\mathcal{R}))\,a_1, \\
    \overline{u}_{i}(\mathcal{R},a)  & \coloneqq a\,w_i(\mathcal{R})+(1-w_i(\mathcal{R}))\,a_L.
  \end{aligned}
\end{equation}
Both endpoints are
attained when all outsiders hold $a_1$ or $a_L$, respectively. The interval narrows toward $a$ as $w_i(\mathcal{R})$ increases (\cref{fig:influence_threshold}). To preserve $a$, a member must resist changes to both lower and higher actions.

Requiring $\underline{u}_i(\mathcal{R},a)\geq\underline{\theta}_i(a)$ and $\overline{u}_i(\mathcal{R},a)\leq\overline{\theta}_i(a)$ and solving for $w_i(\mathcal{R})$
gives two lower bounds on $w_i(\mathcal{R})$, which motivate the following definition.

\begin{definition}[Influence threshold]
  \label{def:support_threshold}
  For an agent $i$ and an action $a$, let
  $\kappa_i(a)\coloneqq\max\{\kappa_i^-(a),\kappa_i^+(a)\}$ be the \mbox{\emph{influence threshold}, where}
  \begin{equation}
    \label{eq:critical}
    \kappa_i^-(a)\coloneqq\frac{\underline{\theta}_i(a)-a_1}{a-a_1},\qquad
    \kappa_i^+(a)\coloneqq\frac{a_L-\overline{\theta}_i(a)}{a_L-a},
  \end{equation}
  with the convention $\kappa_i^-(a_1)=\kappa_i^+(a_L)=0$.
\end{definition}

\begin{samepage}
By \cref{asm:self_consistent}, $\kappa_i(a)\in[0,1]$.
The following theorem gives an exact condition \mbox{for \emph{action preservation}.}

\begin{theorem}[Action preservation]
  \label{thm:retention}
  Let $a\in\mathcal{A}$ and $\emptyset\neq\mathcal{R}\subseteq\mathcal{V}$.
  Then $(\mathcal{R},a)$ is robust if and only if
  \begin{equation}
    \label{eq:robust_condition}
    w_i(\mathcal{R})\geq\kappa_i(a)\quad \text{for every } i \in \mathcal{R}.
  \end{equation}
  Equivalently, $\mathcal{U}_i(\mathcal{R},a)\subseteq\overline{\mathcal{I}}_i(a)$ for every $i\in\mathcal{R}$.
\end{theorem}
\end{samepage}

\begin{proof}
  For each finite endpoint of $\mathcal{I}_i(a)$, \eqref{eq:input_bounds} gives \mbox{the corresponding identity}
  \begin{align*}
    \underline{u}_i(\mathcal{R},a)-\underline{\theta}_i(a)
     & =(a-a_1)\bigl(w_i(\mathcal{R})-\kappa_i^-(a)\bigr), \\
    \overline{\theta}_i(a)-\overline{u}_i(\mathcal{R},a)
     & =(a_L-a)\bigl(w_i(\mathcal{R})-\kappa_i^+(a)\bigr).
  \end{align*}
  The inclusion $\mathcal{U}_i(\mathcal{R},a)\subseteq\overline{\mathcal{I}}_i(a)$ holds if and only if $w_i(\mathcal{R})\geq\kappa_i(a)$, since infinite endpoints impose no restriction.

  If $x_i\in\mathcal{I}_i(a)$ and $u_i\in\overline{\mathcal{I}}_i(a)$, their convex combination in \eqref{eq:dynamics} remains in $\mathcal{I}_i(a)$: since $\alpha_i>0$, the next opinion stays strictly below $\overline{\theta}_i(a)$ whenever that endpoint is finite.

  Conversely, suppose that an input endpoint lies below $\underline{\theta}_i(a)$. This input is attained when all outsiders hold $a_1$, and with $x_i=\underline{\theta}_i(a)$ the next opinion lies below $\mathcal{I}_i(a)$. Similarly, if an input endpoint lies above $\overline{\theta}_i(a)$, it is attained when all outsiders hold $a_L$, and with $x_i=\overline{\theta}_i(a)-\varepsilon$ for sufficiently small $\varepsilon>0$ the next opinion lies above $\mathcal{I}_i(a)$. In both cases, the other members can hold opinion $a$, so $\bm{x}\in\mathcal{X}_a(\mathcal{R})$.
\end{proof}

We call \eqref{eq:robust_condition} the \emph{cohesion condition}. As in~\cite{morris2000contagion}, it requires members to place enough influence weight on the group, but the required weight depends on each member's action rule.

\begin{remark}[Binary CODA]
  \label{rem:binary_coda}
  With actions $\{-1,1\}$ and a common threshold $0$, $\kappa_i(-1)=\kappa_i(1)=1/2$. With equal weights on neighbors, the cohesion condition becomes $|\mathcal{N}_i\cap\mathcal{R}|\geq|\mathcal{N}_i\setminus\mathcal{R}|$ for every $i\in\mathcal{R}$. This coincides with the criterion used for robust polarized clusters in~\cite{chowdhury2016continuous}, although the two models use different opinion updates and \mbox{threshold tie rules.}
\end{remark}

  \begin{remark}[Robust action cluster within a candidate set]
    \label{rem:peeling}
    For a fixed action $a$, unions of robust action clusters are robust by \eqref{eq:robust_condition}. Given a candidate set $\mathcal{S}\subseteq\mathcal{V}$, repeatedly remove agents whose influence weight on the remaining set is below $\kappa_i(a)$. No member of a robust subset of $\mathcal{S}$ can be removed. At termination, the remaining set is therefore either empty or the unique largest robust action cluster contained in $\mathcal{S}$. If none remain, $\mathcal{S}$ has no robust action cluster for $a$.
  \end{remark}

Robust action clusters can also make disagreement persist. If disjoint groups $\mathcal{R}$ and $\mathcal{R}'$ are robust action clusters for actions $a$ and $b$, respectively, with $a\neq b$, every trajectory starting in $\mathcal{X}_a(\mathcal{R})\cap\mathcal{X}_b(\mathcal{R}')$
preserves both actions, and such trajectories never reach action consensus.

% Thus action disagreement persists and consensus is impossible.

We next ask when a robust action cluster makes other agents adopt its action permanently.

\section{Guaranteed Adoption}
\label{sec:propagation}

Fix $a\in\mathcal{A}$ and let $\mathcal{R}_0$ be a robust action cluster for action $a$,
which we call a \emph{robust seed}. A trajectory is \emph{admissible} for this seed and target action $a$ if $\bm{x}(0)\in\mathcal{X}_{a}(\mathcal{R}_0)$.

An agent $i$ \emph{adopts $a$ permanently} if $q_i(x_i(t))=a$
for all sufficiently large $t$. A \emph{guaranteed adopter} adopts $a$ permanently along every admissible trajectory. This guarantee must hold for all seed opinions consistent with action $a$ and all outsider opinions. An agent may therefore adopt $a$ from some initial states without being a guaranteed adopter. A trajectory reaches \emph{permanent $a$-consensus} if all agents hold $a$ \mbox{from some time on.}

We first give a sufficient condition for an outsider to be a guaranteed adopter. Iterating it yields a cascade of guaranteed adopters; when the cascade reaches every agent, every admissible trajectory reaches permanent $a$-consensus.

\subsection{An Adoption Condition}
\label{subsec:recruitment}

The cohesion condition of \cref{thm:retention} concerns agents already holding action $a$. To make an outsider adopt from any initial opinion, it suffices to keep its social input inside $\mathcal{I}_i(a)$
with a positive margin from each finite endpoint. The following lemma expresses this requirement as the \emph{adoption condition} \eqref{eq:strict_input}, a strict version of the cohesion condition.

\begin{lemma}[Adoption condition]
  \label{prop:recruitment}
  Suppose $(\mathcal{R},a)$ is robust. If an outsider $i\notin\mathcal{R}$ satisfies
  \begin{equation}
    \label{eq:strict_input}
    w_i(\mathcal{R})>\kappa_i(a),
  \end{equation}
  or, equivalently, $\mathcal{U}_i(\mathcal{R},a)\subseteq\mathcal{I}_i^\circ(a)$, then there is a finite time $T_i$, independent of the initial state, such that every trajectory from $\mathcal{X}_a(\mathcal{R})$ satisfies $q_i(x_i(t))=a$ for all $t\geq T_i$.
\end{lemma}

\begin{proof}
  The equivalence follows from \eqref{eq:input_bounds} and
  \eqref{eq:critical}. Let
  \(
  \delta_i\coloneqq\min\{\underline{u}_i(\mathcal{R},a)-\underline{\theta}_i(a),\,
  \overline{\theta}_i(a)-\overline{u}_i(\mathcal{R},a)\}>0,
  \)
  omitting the term of an infinite endpoint.
  By robustness, $u_i(t)\in\mathcal{U}_i(\mathcal{R},a)$ for all $t$; iterating \eqref{eq:dynamics} and using $x_i(0)\in[-1,1]$ gives
  \(
  \underline{u}_i(\mathcal{R},a)-2\alpha_i^t\leq x_i(t)
  \leq\overline{u}_i(\mathcal{R},a)+2\alpha_i^t.
  \)
  Since $\alpha_i\in(0,1)$, we may take the finite bound
    \[
      T_i\coloneqq\min\{t\in\mathbb{Z}_{\geq0}:2\alpha_i^t<\delta_i\}.
    \]
    For $t\geq T_i$, $x_i(t)$ lies strictly inside $\mathcal{I}_i(a)$.
\end{proof}

The strict inequality cannot, in general, be relaxed to equality: with a constant input equal to a finite endpoint of $\mathcal{I}_i(a)$, an opinion starting beyond that endpoint converges to it without ever entering $\mathcal{I}_i(a)$. The preservation and adoption conditions are independent of $\alpha_i\in(0,1)$, whereas the adoption-time bound depends on $\alpha_i$.

\subsection{Cascade}
\label{subsec:direct_closure}

An outsider that adopts an action permanently can help other outsiders meet the adoption condition. Iterating this condition yields a cascade of guaranteed adopters.

\begin{definition}[Cascade]
  \label{def:direct_closure}
  Starting from a robust seed $\mathcal{R}_0$, let $\mathcal{R}_{k+1}=\Phi_a(\mathcal{R}_{k})$ for $k=0,1,\dots$, where
  \begin{equation}
    \label{eq:cascade}
    \Phi_a(\mathcal{R})\coloneqq\mathcal{R}\cup\{i\mathrel{\in}\mathcal{V}:w_i(\mathcal{R})>\kappa_i(a)\}.
  \end{equation}
  This increasing sequence stabilizes after at most $N-|\mathcal{R}_0|$ strict enlargements;
  its limit is the \emph{cascade closure} $\operatorname{cl}_a(\mathcal{R}_0)$.
\end{definition}

Each step of the cascade preserves robustness and
guarantees permanent adoption for new members.

\begin{proposition}[{Cascade propagation}]
  \label{prop:cascade}
  Let $\mathcal{R}_0$ be a robust seed for action $a$. Every set $\mathcal{R}_{k}$
  in the cascade sequence of \cref{def:direct_closure} is a robust action cluster for action $a$, and every member of $\operatorname{cl}_a(\mathcal{R}_0)$ is a guaranteed adopter of action $a$.
\end{proposition}

\begin{proof}
  We show by induction on $k$ that $\mathcal{R}_{k}$ is robust and that there is a finite time $\tau_{k}$ such that every admissible trajectory satisfies $\bm{x}(t)\in\mathcal{X}_a(\mathcal{R}_{k})$ for all $t\geq\tau_{k}$. For $k=0$, both claims hold with $\tau_0=0$. Assume they hold for $k$. Each $i\in\mathcal{R}_{k+1}\setminus\mathcal{R}_{k}$ satisfies $w_i(\mathcal{R}_{k})>\kappa_i(a)$. Since the dynamics are time-invariant, \cref{prop:recruitment} applied from time $\tau_{k}$ gives $q_i(x_i(t))=a$ for all $t\geq\tau_{k}+T_i$. The second claim therefore holds for $k+1$, with $\tau_{k+1}\coloneqq\tau_{k}+\max_{i\in\mathcal{R}_{k+1}\setminus\mathcal{R}_{k}}T_i$ (or $\tau_{k+1}\coloneqq\tau_{k}$ if no agent joins). Moreover, $w_i(\mathcal{R}_{k+1})\geq w_i(\mathcal{R}_{k})$ for every $i$, so every member of $\mathcal{R}_{k+1}$ satisfies \eqref{eq:robust_condition}: new members because $w_i(\mathcal{R}_{k})>\kappa_i(a)$, and members of $\mathcal{R}_{k}$ by \cref{thm:retention}. Hence $\mathcal{R}_{k+1}$ is robust by \cref{thm:retention}. Applying the second claim to the limit $\operatorname{cl}_a(\mathcal{R}_0)$ shows that each of its members is a guaranteed adopter.
\end{proof}

Thus $\operatorname{cl}_a(\mathcal{R}_0)=\mathcal{V}$
is sufficient for permanent $a$-consensus along every admissible trajectory.

The iteration in \cref{def:direct_closure} is a progressive linear-threshold cascade with thresholds $\kappa_i(a)$ and the strict adoption rule \eqref{eq:strict_input}. As in~\cite{morris2000contagion}, cohesive groups can block it. A nonempty group $\mathcal{B}$ is an \emph{obstruction} if $w_i(\mathcal{B})\geq1-\kappa_i(a)$, or equivalently $w_i(\mathcal{V}\setminus\mathcal{B})\leq\kappa_i(a)$, for all $i\in\mathcal{B}$. Then no member of $\mathcal{B}$ satisfies \eqref{eq:strict_input} for $\mathcal{R}=\mathcal{V}\setminus\mathcal{B}$, even when every other agent has adopted $a$. The cascade reaches every agent if and only if every obstruction intersects the seed. An obstruction blocks only the cascade, not adoption itself: agents outside $\operatorname{cl}_a(\mathcal{R}_0)$ may still adopt $a$ along particular trajectories. For $L=2$, however, the cascade is exact. If $b$ denotes the other action, then $\kappa_i(b)=1-\kappa_i(a)$, so every obstruction $\mathcal{B}$ disjoint from the seed is a robust action cluster for $b$, and its members keep $b$ along every admissible trajectory starting in $\mathcal{X}_b(\mathcal{B})$. Since $\mathcal{V}\setminus\operatorname{cl}_a(\mathcal{R}_0)$ is such an obstruction whenever it is nonempty, $\operatorname{cl}_a(\mathcal{R}_0)$ is then exactly the set of guaranteed adopters. For $L>2$,
the cascade gives only a sufficient condition for guaranteed adoption, since the worst-case inputs of different members of an obstruction need not be simultaneously realizable.

\section{Numerical Illustrations}
\label{sec:numerics}

We illustrate \cref{thm:retention} and the cascade of \cref{def:direct_closure} on two networks. Each example compares two runs: one with heterogeneous action rules, and one in which all agents
use the \emph{homogeneous rule}. Each threshold of this common rule is the mean, over all agents, of the corresponding heterogeneous thresholds. Within each example, the graph and initial opinions are identical in both runs. \Cref{tab:rules} lists the parameters of both examples, including the thresholds of every action rule.
We say that the action profile remains fixed after time $t_0$ if no action changes after $t_0$. We verify this with the following criterion: every social input at time $t_0$ lies in the closure of the agent's current action interval. Since $\alpha_i>0$, the update \eqref{eq:dynamics} then keeps every opinion in its action interval, so neither the actions nor the inputs change.

\begin{table}[ht]
  \centering
  \caption{Parameters of \cref{sec:numerics}. Both examples use undirected graphs with uniform weights $W_{ij}=1/|\mathcal{N}_i|$ for $j\in\mathcal{N}_i$ and inertia $\alpha_i=0.8$. Each action rule is given by its thresholds $(\theta_{i,1},\dots,\theta_{i,L-1})$ and the influence threshold $\kappa_i(a)$ of \eqref{eq:critical} for the target action $a$.}
  \label{tab:rules}
  \footnotesize
  \setlength{\tabcolsep}{4pt}
  \begin{tabular}{@{}llcc@{}}
    \toprule
    Rule (agents)      & Action thresholds                 & $a$          & $\kappa_i(a)$    \\
    \midrule
    \multicolumn{4}{@{}l}{\cref{subsec:cascade_example}: $\mathcal{A}=\{-1,-1/2,1/2,1\}$}    \\
    A (8)              & $(-0.70,\,-0.30,\,0.80)$          & $1/2$        & $0.47$   \\
    B (3)              & $(-0.70,\,0,\,0.80)$              & $1/2$        & $0.67$   \\
    Homogeneous (11)   & $(-0.70,\,-0.218,\,0.80)$ & $1/2$        & $0.52$   \\
    \midrule
    \multicolumn{4}{@{}l}{\cref{subsec:grid_clusters}: $\mathcal{A}=\{-1,0,1\}$}             \\
    C (1250)           & $(-0.60,\,0.60)$                  & $0$ / $\pm1$ & $0.40$ / $0.80$  \\
    D (1250)           & $(-0.10,\,0.10)$                  & $0$ / $\pm1$ & $0.90$ / $0.55$  \\
    Homogeneous (2500) & $(-0.35,\,0.35)$                  & $0$ / $\pm1$ & $0.65$ / $0.675$ \\
    \bottomrule
  \end{tabular}
\end{table}

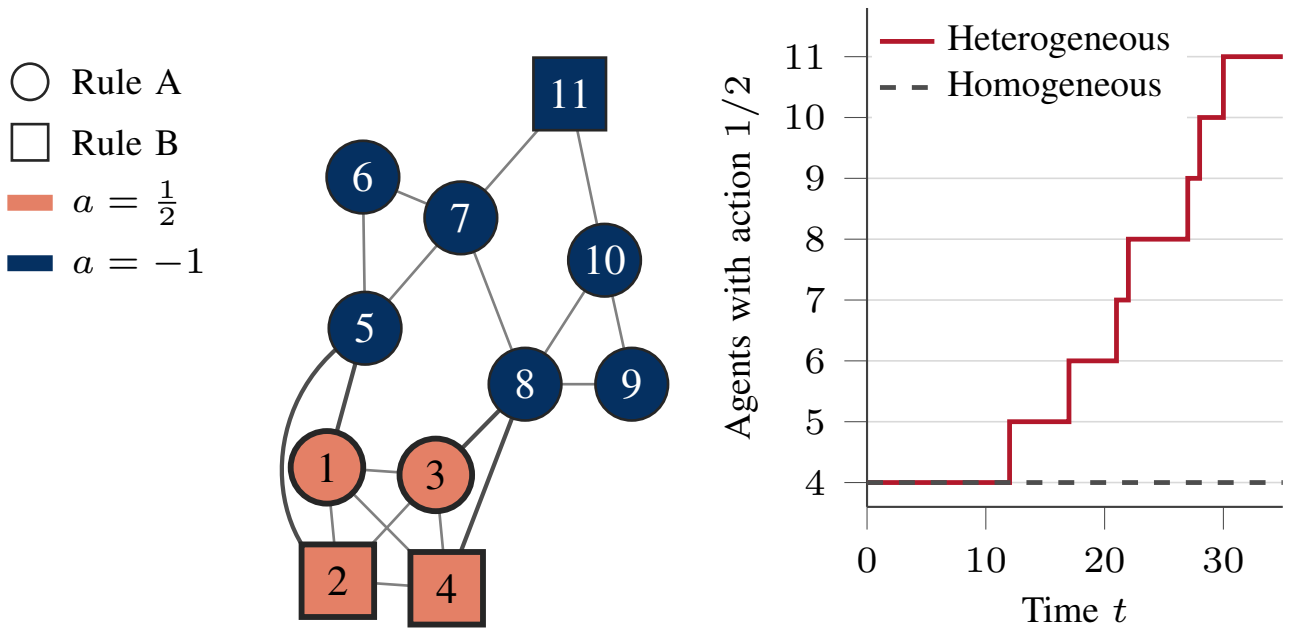
\begin{figure}[htbp]
  \centering
  \begin{subfigure}[b]{0.53\columnwidth}
    \centering
    % Fig. cascade (a): eleven-agent network with heterogeneous rules in the seed.
% Coordinates are drawing choices only; all 19 undirected edges are preserved.
\definecolor{cascadeNegative}{HTML}{053061}%
\definecolor{cascadeInitial}{HTML}{E48066}%
\begin{tikzpicture}[
    font=\scriptsize,
    agent/.style={draw=black!85, line width=0.5pt, minimum size=4.8mm,
      inner sep=0pt, font=\footnotesize},
    ruleA/.style={agent,circle},
    ruleB/.style={agent,rectangle},
    seed/.style={fill=cascadeInitial,text=black,line width=1.1pt},
    negative/.style={fill=cascadeNegative,text=white},
    edge/.style={draw=black!50,line width=0.5pt},
    seedlink/.style={draw=black!70,line width=0.8pt},
  ]
  % Network, compressed horizontally; node shapes and labels keep their size.
  \begin{scope}[xshift=1.8cm,xscale=0.72]
  \coordinate (p1) at (0.45,2.00);
  \coordinate (p2) at (0.55,1.25);
  \coordinate (p3) at (1.45,1.95);
  \coordinate (p4) at (1.55,1.20);
  \coordinate (p5) at (0.80,2.92);
  \coordinate (p6) at (0.78,3.92);
  \coordinate (p7) at (1.68,3.65);
  \coordinate (p8) at (2.27,2.55);
  \coordinate (p9) at (3.25,2.55);
  \coordinate (p10) at (3.00,3.37);
  \coordinate (p11) at (2.68,4.47);
  \draw[edge] (p1)--(p2)--(p4)--(p3)--cycle;
  \draw[edge] (p1)--(p4) (p2)--(p3);
  \draw[edge] (p5)--(p6)--(p7)--cycle;
  \draw[edge] (p8)--(p9)--(p10)--cycle;
  \draw[edge] (p7)--(p8) (p7)--(p11)--(p10);
  \draw[seedlink] (p5)--(p1) (p8)--(p3) (p8)--(p4);
  % Route 5--2 outside agent 1 so the two links to the seed are distinct.
  \draw[seedlink] (p5) .. controls (-0.20,2.50) and (-0.15,1.55) .. (p2);
  \node[ruleA,seed] (n1) at (p1) {1};
  \node[ruleB,seed] (n2) at (p2) {2};
  \node[ruleA,seed] (n3) at (p3) {3};
  \node[ruleB,seed] (n4) at (p4) {4};
  \node[ruleA,negative] (n5) at (p5) {5};
  \node[ruleA,negative] (n6) at (p6) {6};
  \node[ruleA,negative] (n7) at (p7) {7};
  \node[ruleA,negative] (n8) at (p8) {8};
  \node[ruleA,negative] (n9) at (p9) {9};
  \node[ruleA,negative] (n10) at (p10) {10};
  \node[ruleB,negative] (n11) at (p11) {11};
  \end{scope}

  % Legend in the upper-left corner, top aligned with the network:
  % shape gives the rule, a swatch the initial action; thick borders mark the seed.
  \begin{scope}[shift={(0,4.55)}]
  \node[circle,draw=black!85,line width=0.5pt,minimum size=2.4mm,inner sep=0pt]
    at (0.16,0) {};
  \node[anchor=west,inner sep=0pt] at (0.43,0) {Rule A};
  \node[rectangle,draw=black!85,line width=0.5pt,minimum size=2.4mm,inner sep=0pt]
    at (0.16,-0.40) {};
  \node[anchor=west,inner sep=0pt] at (0.43,-0.40) {Rule B};
  \draw[cascadeInitial,line width=3pt] (0.01,-0.80)--(0.31,-0.80);
  \node[anchor=west,inner sep=0pt] at (0.43,-0.80) {$a=\frac{1}{2}$};
  \draw[cascadeNegative,line width=3pt] (0.01,-1.20)--(0.31,-1.20);
  \node[anchor=west,inner sep=0pt] at (0.43,-1.20) {$a=-1$};
  \end{scope}
\end{tikzpicture}%
    \caption{Initial network}
    \label{fig:cascade_network}
  \end{subfigure}%
  \hfill
  \begin{subfigure}[b]{0.45\columnwidth}
    \centering
    % Fig. cascade (b): number of agents holding action 1/2, held constant
% between integer time steps. Data: figures/data/ecc_V_A_eleven_agent.dat
\definecolor{cascadeCurve}{HTML}{B2182B}%
\begin{tikzpicture}[font=\scriptsize]
  \begin{axis}[
      scale only axis, width=2.75cm, height=3.30cm,
      axis lines*=left,
      xmin=0, xmax=35, ymin=3.6, ymax=11.8,
      xtick={0,10,20,30}, ytick={4,5,...,11},
      xlabel={Time $t$}, ylabel={Agents with action $1/2$},
      tick label style={font=\scriptsize}, label style={font=\scriptsize},
      xlabel style={inner sep=1pt}, ylabel style={inner sep=1pt},
      tick align=outside, tick style={black!70}, axis line style={black!75},
      ymajorgrids, grid style={black!15,line width=0.3pt},
      % Legend in the empty upper-left region, above both curves
      % The y-range starts just below the seed size 4 so the mean-rule line stays
      % visible above the x-axis; ymax leaves headroom for the legend.
      legend style={font=\scriptsize,draw=none,fill=white,
        at={(0.01,1.0)},anchor=north west,inner sep=1pt,
        cells={anchor=west},row sep=-1.5pt},
      legend image post style={xscale=0.6},
      clip=true,
    ]
    \addplot[cascadeCurve,line width=0.9pt,const plot,no marks]
      table[x=t,y=heterogeneous] {figures/data/ecc_V_A_eleven_agent.dat};
    \addlegendentry{Heterogeneous}
    \addplot[black!70,line width=0.9pt,dashed,const plot,no marks]
      table[x=t,y=mean] {figures/data/ecc_V_A_eleven_agent.dat};
    \addlegendentry{Homogeneous}
  \end{axis}
\end{tikzpicture}%
    \caption{Agents holding action $1/2$}
    \label{fig:cascade_counts}
  \end{subfigure}
  \caption{Propagation of action $1/2$ from a robust minority of four among eleven agents. (a) Initial network: shapes denote rules A and B in the heterogeneous run, colors give initial actions, and thick borders mark the seed $\mathcal{R}_0=\{1,2,3,4\}$, which contains both rules. In the other run, every agent uses the homogeneous rule. (b) Number of agents currently holding action $1/2$, from identical initial opinions. Under heterogeneous rules all eleven adopt permanently by $t=30$; under the homogeneous rule only the four seed members hold action $1/2$. Only $0\leq t\leq35$ is displayed; both runs continue to $T=180$.}
  \label{fig:cascade}\vspace{-3mm}
\end{figure}

\begin{figure*}[t]
  \centering
  \includegraphics[width=\textwidth]{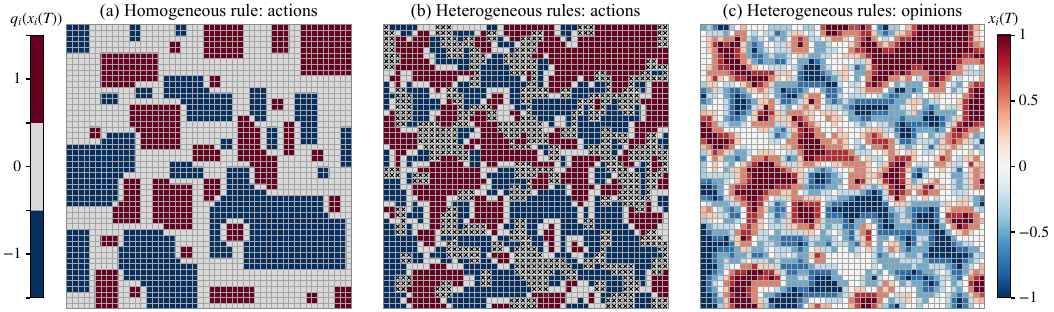}
  \caption{Robust action clusters on an open (non-periodic) $50\times50$ lattice with actions $\{-1,0,1\}$, from the same uniform random initial opinions; both final action profiles remain fixed.
    (a) Final actions $q_i(x_i(T))$, at $T=1200$, when every agent uses the homogeneous rule: the final action profile contains no robust action cluster.
    (b) Final actions under heterogeneous rules: half of the agents, placed at random, use rule~C and the others rule~D (\cref{tab:rules}); crosses mark the agents in robust action clusters, all holding action $0$.
    (c) Final private opinions $x_i(T)$ under heterogeneous rules.}
  \label{fig:grid_clusters}
\end{figure*}

\subsection{Propagation of an Intermediate Action}
\label{subsec:cascade_example}

The first example shows that mean thresholds alone do not determine whether an intermediate action propagates from a robust seed.
The action set is $\mathcal{A}=\{-1,-1/2,1/2,1\}$, and the network of \cref{fig:cascade}(a) has eleven agents. Agents $1$ to $4$ form a complete graph and constitute the seed $\mathcal{R}_0$. The seed starts with action $1/2$ and the seven outsiders with action $-1$. In the heterogeneous run, rules~A and~B are assigned as shown in \cref{fig:cascade}(a), with both rules represented in the seed.

Every seed member places influence weight $3/4$ on $\mathcal{R}_0$, so its worst-case input interval $\mathcal{U}_i(\mathcal{R}_0,1/2)=[1/8,5/8]$, given by \eqref{eq:input_bounds}, lies strictly inside its action interval for action $1/2$ under every rule. Hence $\mathcal{R}_0$ is a robust seed for action $1/2$ in both runs, by \cref{thm:retention}.
Under heterogeneous rules, outsider~5 places weight $1/2$ on the seed, above rule~A's influence threshold of approximately $0.47$, and joins the cascade first. Each later outsider places at least $1/2$ on the growing group when it joins, and outsider~11, the only outsider using rule~B, joins last with weight $1>0.67$. Thus $\operatorname{cl}_{\frac{1}{2}}(\mathcal{R}_0)
  =\mathcal{V}$, and every agent is a guaranteed adopter of action $1/2$.
Under the homogeneous rule, the only outsiders with seed neighbors, agents~5 and~8, place weights $1/2$ and $2/5$ on the seed, both below $\kappa_i(1/2)\approx0.52$, so \mbox{the cascade stops at $\mathcal{R}_0$.}

\Cref{fig:cascade}(b)
shows the two runs. Under heterogeneous rules, all outsiders permanently adopt action $1/2$. Agent~5 starts this propagation: its initial input $-1/4$ lies above rule~A's middle threshold $-0.30$ but below the mean middle threshold of approximately $-0.218$. Under the homogeneous rule, agents~5 and~8 move from $-1$ to $-1/2$ and keep inputs $-1/4$ and $-2/5$, below the mean middle threshold, while all other outsiders remain at $-1$. The resulting action profile remains fixed, with only the seed holding action $1/2$.
Indeed, every social input lies strictly inside its agent's current action interval, so the criterion stated at the beginning of this section applies. The stalled cascade alone would show only that the cascade criterion does not certify adoption. The seed preserves its action in both runs, but replacing the individual rules by their mean stops the propagation.

\subsection{Robust Action Clusters on a Square Lattice}
\label{subsec:grid_clusters}

The second example
asks which groups in a fixed action profile are robust action clusters.
A fixed action profile does not establish robustness. \Cref{thm:retention} identifies, from the final action profile alone, the groups that would keep their actions whatever the other agents do. With two actions, every agent of a fixed action profile belongs to such a group. Indeed, let $\mathcal{R}$ be the set of agents holding $a_L$. Every other agent holds $a_1$, so each member of $\mathcal{R}$ receives the constant input $\underline{u}_i(\mathcal{R},a_L)$ of \eqref{eq:input_bounds} and keeps $a_L$ as its opinion approaches that input. This input therefore lies in the closure of the action interval, and $(\mathcal{R},a_L)$ is robust by \cref{thm:retention}. The same holds for $a_1$. This is the situation of the binary lattice in~\cite{chowdhury2016continuous}. We therefore use three actions,
the smallest action set in which a fixed action profile need not consist of robust action clusters.

Following~\cite{chowdhury2016continuous}, we use an open $50\times50$ square lattice with horizontal and vertical edges and actions $\mathcal{A}=\{-1,0,1\}$. Initial opinions are independent uniform draws from $[-1,1]$.

In one run, every agent uses the homogeneous rule; in the other, rules~C and~D are assigned at random to equal numbers of agents (\cref{tab:rules}).
On the lattice, an interior agent, away from the boundary, has four neighbors, so an influence threshold translates into a number of neighbors. Under the homogeneous rule, $\kappa_i(0)=0.65$ and $\kappa_i(\pm1)=0.675$,
so an interior agent needs three of its four neighbors in its group, whatever action it holds. Under rule~C, $\kappa_i(0)=0.4$,
so an interior agent holding action $0$ needs only two; under rule~D, $\kappa_i(0)=0.9$, so it needs all four.

In any group disjoint from the lattice boundary, the leftmost member of the top row has its upper and left neighbors outside the group, so it places influence weight at most $1/2$ on the group.
In \cref{tab:rules}, all influence thresholds for actions $\pm1$ exceed $1/2$, and only rule~C has $\kappa_i(0)\leq1/2$.
By \cref{thm:retention}, no group disjoint from the boundary is a
robust action cluster under the homogeneous rule.
Under heterogeneous rules, among groups disjoint from the boundary, only those holding action $0$ can be robust. \Cref{fig:grid_clusters} shows both consequences. Under the homogeneous rule, the final action profile remains fixed but contains no robust action cluster, including at the boundary (\cref{fig:grid_clusters}(a)). The absence of robust clusters is also observed in $15$ additional initializations.

% Under heterogeneous rules, the final action profile also remains fixed. The largest robust subset of the final zero-action set contains $701$ agents (\cref{fig:grid_clusters}(b)). We apply the procedure of \cref{rem:peeling} to each final action set and split the remaining subsets into connected robust action clusters.

Under heterogeneous rules, the final action profile also remains fixed. Applying the procedure of \cref{rem:peeling} to each final action set leaves no agents for actions $\pm1$ and 701 agents for action $0$ (\cref{fig:grid_clusters}(b)); we split them into connected robust action clusters.

The other remaining agents also keep their actions, but none belongs to a robust action cluster. Their actions remain fixed only because of the actions currently held around them. For each connected group of these agents holding a common action, we start from the final state and fix all outsider actions to $-1$ in one run and to $1$ in another. At least one run causes an agent in the group to change its action. Every robust action cluster keeps action $0$ in both runs, as \cref{thm:retention} predicts. Private opinions
do not reveal this distinction. Agents holding action $0$ have opinions ranging over $[-0.5,0.5]$ whether or not they belong to a robust action cluster (\cref{fig:grid_clusters}(b)--(c)). As in \cref{subsec:cascade_example}, the homogeneous rule
does not reproduce the behavior under heterogeneous rules. Here, it excludes the robust action clusters that rule~C makes possible.

\section{Conclusion}
\label{sec:conclusion}

We characterized robust action clusters in networks with continuous private opinions and heterogeneous action rules. A group preserves a common action regardless of outsiders' actions if and only if each member satisfies an action-dependent cohesion condition. A strict version yields a cascade of guaranteed adopters, which identifies all guaranteed adopters in the binary case. Numerical examples show that replacing heterogeneous thresholds by their means can alter both action propagation and the presence of robust clusters. Future work includes addressing seed formation, adoption through intermediate actions beyond the cascade, and asynchronous updates.

\balance
\bibliographystyle{IEEEtran}
\bibliography{biblio2}

\end{document}